\documentclass[submission,copyright,creativecommons]{eptcs}

\providecommand{\event}{Agents and Robots for reliable Engineered Autonomy (AREA 2020)}

\usepackage{breakurl}

\usepackage{algorithm} 
\usepackage{algpseudocode}
\usepackage{graphicx}
\usepackage{multicol}
\usepackage{listings}
\usepackage{xcolor}
\usepackage{xspace}
\usepackage{url}

\hypersetup{hidelinks}

\algnewcommand\In{\textbf{in}}
\algnewcommand\Is{\textbf{is}}
\algnewcommand\Not{\textbf{not}}

\newcommand\comparison[3]%
{%
	\begin{minipage}[t]{.48\textwidth}
		\lstinputlisting[language=Java,basicstyle=\ttfamily\tiny]{#1}
	\end{minipage}
	\hfill
	\begin{minipage}[t]{.48\textwidth}
		\lstinputlisting[language=Java,basicstyle=\ttfamily\tiny]{#2}
		\lstinputlisting[language=Java,basicstyle=\ttfamily\tiny]{#3}
	\end{minipage}
}

\lstdefinelanguage{jadescript}{
	morekeywords = { agent, behaviour, ontology, concept, predicate, proposition, uses,
	for, extends, do, on, when, is, message, send, content, cyclic, oneshot, of,
	text, integer, aid, boolean, or, and, as, create, destroy, activate,
	performative, receivers, with, not, in, log, new, true, false, action, double,
	float, procedure, function, return, break, continue, to, one, shot, if, else,
	property, request, inform, matches, deactivate, module, percept, invoke, any },
	sensitive = true,
	morecomment = [l]{\#},
	morecomment = [s]{/*}{*/},
	morestring = [b]"
}

\newcommand{\JADE}{JADE\xspace}
\newcommand{\Jadescript}{Jadescript\xspace}

\begin{document}

\title{Exploratory Experiments on Programming\\Autonomous Robots in Jadescript}

\author{
Eleonora Iotti
\quad
Giuseppe Petrosino
\quad
Stefania Monica
\quad
Federico Bergenti
\institute{Dipartimento di Scienze Matematiche, Fisiche e Informatiche\\
Universit\`a degli Studi di Parma, 43124 Parma, Italy}
\email{\{eleonora.iotti,stefania.monica,federico.bergenti\}@unipr.it}
\email{giuseppe.petrosino@studenti.unipr.it}
}

\def\titlerunning{Exploratory Experiments on Programming Autonomous Robots in Jadescript}
\def\authorrunning{E. Iotti, G. Petrosino, S. Monica \& F. Bergenti}

\maketitle

\begin{abstract}
This paper describes exploratory experiments to validate the possibility of programming autonomous robots using an agent-oriented programming language. Proper perception of the environment, by means of various types of sensors, and timely reaction to external events, by means of effective actuators, are essential to provide robots with a sufficient level of autonomy. The agent-oriented programming paradigm is relevant with this respect because it offers language-level abstractions to process events and to command actuators. A recent agent-oriented programming language called \Jadescript is presented in this paper together with its new features specifically designed to handle events. Exploratory experiments on a simple case-study application are presented to show the validity of the proposed approach and to exemplify the use of the language to program autonomous robots.
\end{abstract}

\section{Introduction}
\label{sec:introduction}
\emph{AOP} (\emph{Agent-Oriented Programming})~\cite{aop:shoham1997} is a programming paradigm that aims at providing effective languages and tools to develop agent-based software systems (e.g.,~\cite{aop:bradshaw1997}). The most challenging parts of the development of complex agent-based software systems are expected to be leveraged by AOP languages and tools. AOP languages and tools allow programmers to reason on high-level views of multi-agent systems instead of concentrating on fine-grained details that tend to distract attention from targeted problems. Examples of such challenging parts of the development of complex agent-based software systems are the deployment of agents to network hosts, the routing of messages across the network, and the handling of sensor information. The abstractions that AOP languages normally support and that make such languages suitable for agent-based software development are discussed, for example, in the specifications from \emph{FIPA} (\emph{Foundation for Intelligent Physical Agents}), now IEEE FIPA Standards Committee (\url{http://www.fipa.org}), and in the literature on \emph{AOSE} (\emph{Agent-Oriented Software Engineering}) (e.g.,~\cite{aose2004}). The effectiveness of such abstractions in the development of complex agent-based software systems has progressively increased the interest in AOP for industrial and academic uses.

\Jadescript~\cite{jadescript:agere2018} is briefly presented in Section~\ref{sec:jadescript_features} as an AOP language based on \emph{JADE} (\emph{Java Agent DEvelopment framework})~(e.g.,~\cite{jade:book2007,jade:story}) whose main objectives are to ease the use of JADE and to help programmers develop complex agent-based software systems at a high level of abstraction. Besides the presentation of \Jadescript, the main contribution of this paper is to show how an AOP language like \Jadescript can be used to program autonomous robots. A case-study application, which calls for the management of sensors, the handling of events, and the use of actuators, is discussed in Section~\ref{sec:example}. Therefore, in Section~\ref{sec:event_handling}, a detailed description of a new \Jadescript feature designed to process percepts and to make them available to agents is presented. Such a new feature is based on the native support for pattern matching that has been recently added to the language~\cite{jadescript:woa2019}, as recalled in Section~\ref{sec:event_handling}. Programmers use the pattern-matching support that the language natively provides to clearly state the characteristics of the managed percepts, their intended scope, and their priorities. Note that the recent introduction of the pattern-matching support is a substantial step forward with respect to the early versions of the language (e.g.,~\cite{jadescript:woa2018,jadescript:aixia2018}) because it extends the applicability of the language to better include event-driven programming and its applications.

The remaining of this paper is organized as follows. The major progresses of AOP are summarized in Section~\ref{sec:related_work} in terms of a brief description of some of the most important AOP languages. A brief overview of \Jadescript and of its current features is presented in Section~\ref{sec:jadescript_features}. The most recent improvements of the language, which are intended to better support robotic applications, are detailed in Section~\ref{sec:event_handling}. The case-study application designed to validate such improvements is described and commented in Section~\ref{sec:example}. Finally, a short discussion on the proposed approach to AOP concludes the paper.

\section{Related Work}
\label{sec:related_work}
\Jadescript is closely related to \JADE, and it provides practical support to programmers to use some of the most relevant abstractions that \JADE offers, namely agents, behaviours~\cite{jade:bellifemine2005}, and (communication) ontologies~\cite{jade:ontology2005}. \Jadescript agents are fully interoperable with \JADE agents because the ultimate goal of \Jadescript is to reduce the complexity of building agent-based software systems that use \JADE as their underlying infrastructure. As stated in a recent survey~\cite{survey:kravari2015}, \JADE is still one of the most popular agent platforms, even if its first prototype is more than twenty years old~\cite{jade:story}. One of the major reasons for the longevity of \JADE is probably related to its reliability and robustness, which make it suitable for industrial applications designed to take advantage of the peculiar characteristics of software agents in terms of reusability and composability (e.g.,~\cite{esaw2002}). 

\emph{JADEL} (\emph{JADE Language})~\cite{jadel:bergenti2014,jadel:comlan2017} is the direct predecessor of \Jadescript, and it is an AOP language that was designed to ease the use of \JADE. Differently from \Jadescript, JADEL is based on Xtend~\cite{xtext:bettini2013}, which is a dialect of Java integrated with the Xtext infrastructure~\cite{xtext:bettini2013}. The choice of grounding JADEL on Xtend has a direct impact on the language, and the major reasons for the proposal of \Jadescript are to remove such a dependence and to have the freedom to experiment novel linguistic constructs.

Domain-specific languages for the development of agents and multi-agent systems are, for example, \emph{SEA\_ML} (\emph{Semantic web-Enabled Agent Modeling Language})~\cite{seal:demirkol2013}, \emph{CLAIM} (\emph{A Computational Language for Autonomous Intelligent and Mobile Agents})~\cite{claim:2003}, and \emph{3APL} (\emph{An Abstract Agent Programming Language})~\cite{3apl:hindriks1999}. Programming languages with a dedicated support for the reactive processing of events in the scope of robotic applications have also been proposed when AOP was used in the research on robotics, like, for example, in the \emph{PROFETA} (\emph{Python RObotic Framework for dEsigning
sTrAtegies})~\cite{profeta:2017} project. PROFETA is a tool that was developed to program autonomous robots in a declarative way by adopting a \emph{BDI} (\emph{Belief-Desire-Intention}) approach (e.g.,~\cite{agentspeak:rao1996}). PROFETA provides each agent with a list of reactive plans, each of which consists of an event, a condition, and an action. Similarly, \Jadescript provides agents with the event handlers described in Section~\ref{sec:event_handling}. However, PROFETA is inspired by AgentSpeak(L)~\cite{agentspeak:rao1996}, which makes it substantially different from \Jadescript.

According to recent surveys~\cite{survey:buadicua2011,survey:bordini2006}, AOP languages can be classified into declarative or imperative, BDI or procedural, highly specific or general purpose. AgentSpeak(L) and its Java implementation, Jason~\cite{jason:bordini2007}, are examples of declarative BDI languages, while SARL~\cite{sarl:rodriguez2014} is an example of a procedural language that provides the abstractions needed to implement holonic agents and related multi-agent systems. Such languages have their specific ranges of applicability, and a detailed comparison with Jadescript is not among the objectives of this paper. However, both SARL and Jason deal with the sensing of external events, and they can be used to program autonomous robots, as follows.

To the best of our knowledge, SARL  has not yet been tested in robotic applications, although it is clearly possible to use it to control autonomous robots. SARL provides programmers with the possibility of defining the types of application-specific events that the agents can capture. Jadescript does not explicitly provide such a possibility, but it allows programmers to define percepts as particular predicates in the scope of an ontology. Such percepts are then handled by behaviours, as described in Section~\ref{sec:event_handling}. 

Jason beliefs can be used to represent percepts, to the extent that a belief is asserted when a sensor of the agent acquires the pertinent information. Nonetheless, the approach to event handling that Jason provides is known to have relevant concurrency problems~\cite{jasonpitfalls}, which complicates the use of Jason for robotic applications. On the contrary, Jadescript event handlers completely rely on the inherent synchronization that \JADE ensures, and concurrency problems are solved using a priority queue together with the well-known behaviour scheduling policy that characterizes JADE agents, as discussed in Section~\ref{sec:event_handling}.

Besides AOP languages, agent frameworks are also very popular, and many agent-based software systems are written in general-purpose programming languages using such frameworks. \JADE~\cite{jade:bellifemine2005}, JACK~\cite{jack:winikoff2005}, and many other agent frameworks (e.g.,~\cite{survey:kravari2015}) are available to allow the effective use of general-purpose programming languages to develop and deploy agent-based software systems.

\section{Overview of Jadescript}
\label{sec:jadescript_features}
\Jadescript~\cite{jadescript:agere2018} is a novel AOP language designed to allow programmers to adopt agents and multi-agent systems to simplify the development of complex distributed systems. This is achieved by some design choices that make \Jadescript source codes easy to read, but at the same time concise, and similar to pseudocodes in many aspects. Among such design choices, some of the most important are the adoption of semantically-relevant indentation, and the introduction of the \texttt{of} notation, which resembles the common way to refer to the properties of objects in English. Such characteristics of the language, combined with other features like native collection types, make \Jadescript similar to modern scripting languages like Python. \Jadescript agents are nothing but \JADE agents in order to take advantage of the relevant features of the popular agent framework. This is the reason why \Jadescript agents are assumed to be executed by the Java virtual machine, and \Jadescript source codes are intended to be directly compiled to Java. 

\Jadescript is a statically-typed language because the types of all the elements of a source code are known at compile time. The type system of the language is based on five categories of types, namely primitive types, collection types, ontology types, agent types, and behaviour types. Primitive types are \texttt{boolean}, \texttt{integer}, \texttt{float}, \texttt{double}, and \texttt{text}. Collection types are the types of \texttt{list}s and \texttt{map}s. The declaration of structured types is supported in \Jadescript in terms of ontology schemas~\cite{jade:book2007}. \Jadescript borrows from \JADE four categories of ontology schemas, namely predicates, propositions, concepts, and actions. Predicates represent facts in the form of relationships among objects of the world, while propositions represent atomic facts. Concepts represent the objects of the world and their properties, while actions are special concepts that represent the tasks that can be performed by the agents in the multi-agent system. Predicates, concepts, and actions, but not propositions, can have zero or more named properties. The subtyping relationship between ontology schemas can be established to allow derived schemas to be defined by adding and overriding properties of base schemas. Schemas are collected in (communication) ontologies, which are sets of schemas that can be shared among agents to enable communication and to allow agents to cooperate by sharing knowledge expressed in terms of data in agreed formats. Ontologies can extend other ontologies by adding schemas to base ontologies. Figure~\ref{fig:example_onto} shows an example of a simple \Jadescript ontology that comprises two concepts, one action, and one predicate to describe a world made of shapes that agents manipulate.

\begin{figure}[t]
\begin{lstlisting}
ontology Shapes
	concept position(x as double, y as double)

	concept shape(p as position, area as double)

	action createShape(p as position)

	predicate shapeCreated(s as shape)
\end{lstlisting}
\caption{Example of a \Jadescript ontology with two concepts, one action, and one predicate to describe a world made of shapes that agents manipulate.}
\label{fig:example_onto}
\end{figure}

The agents that execute in a \Jadescript multi-agent system are declared with the \texttt{agent} construct, in which properties, functions, procedures, and lifecycle event handlers are defined. An agent type can extend another agent type to inherit, and possibly override, internal declarations from the base agent type. The tasks that agents perform in a \Jadescript multi-agent system are described using behaviours, which are abstractions that \Jadescript borrows from \JADE (e.g.,~\cite{jade:book2007}). The \texttt{behaviour} construct groups three kinds of declarations, namely properties, actions, and event handlers. Actions are portions of procedural code that are executed every time behaviours are scheduled. Event handlers are described in Section~\ref{sec:event_handling}. Just like agent types, a behaviour type can be derived from another behaviour type.

Despite being a statically-typed language, some of the burdens associated with type declarations are relieved from programmers because the \Jadescript compiler supports a limited form of type inference. The compiler is able to automatically and implicitly detect the types of properties and local variables by computing the types of their mandatory initialization expressions. Therefore, the assignment of initialization values is sufficient to declare properties and local variables in \Jadescript without explicitly stating their types in the source code.

\Jadescript agents execute in platforms that are subdivided into containers, and each container is related to a process running on a host of the network. A platform associates each agent with a unique \emph{AID} (\emph{Agent IDentifiers}) that can be used to support communication. In \Jadescript, agents refer to other agents by means of the \texttt{aid} concept, which is also used to send messages with the \texttt{send} construct. The \emph{ACL} (\emph{Agent Communication Language}) specified by the IEEE FIPA Standards Committee is used in \Jadescript to support the communication among agents, and therefore each message is characterized by a performative, which is the name of the communicative act that the sender intends to perform by sending the message, an envelope, and a content. 

\Jadescript allows describing the tasks that agents perform in terms of behaviours. A behaviour represents a fragment of the way an agent acts to bring about a goal, and it is characterized by an internal state and an action. An agent can dynamically engage multiple behaviours at the same time, and all such behaviours are kept in an internal list of active behaviours. Behaviours are executed in the single thread of the agent by means of the cooperative scheduler that characterizes \JADE agents. If no behaviours are active, the agent makes a transition to the waiting state, and it quietly waits for new events to occur. Two kinds of behaviours can be defined in \Jadescript, namely \texttt{cyclic} behaviours and \texttt{one-shot} behaviours. When a \texttt{cyclic} behaviour is active, it is rescheduled multiple times until it is explicitly deactivated. One common use of such behaviours is to implement reactions to changes in the state of the agent or in the state of the world. On the contrary, \texttt{one-shot} behaviours are automatically deactivated at the end of their execution, and they need to be explicitly reactivated if needed. 

Behaviour types are defined by means of the \texttt{behaviour} construct. Behaviours can be constrained to  work with a specific ontology using the \texttt{uses-ontology} clause.
Similarly, they can be constrained to be activated only by the agents of the type declared using the \texttt{for-agent} clause.  Note that if a behaviour is constrained to work on a specific type of agent, it can implicitly use the ontologies that the chosen type of agent references in the \texttt{uses-ontology} clause.

Figure~\ref{fig:example_agent} shows an example of a simple \Jadescript agent with one property and one \texttt{on-create} handler that defines the procedure to be executed upon creation in terms of two behaviours. The two behaviours used by the initialization procedure are shown in Figure~\ref{fig:example_behaviour}. A \texttt{RequestNewShape} behaviour is a \texttt{one-shot} behaviour that, when scheduled for execution, sends a \texttt{request} message to the targeted shape-provider agent. The targeted shape-provider agent, upon successful creation of the requested shape, replies with an \texttt{inform} message. A \texttt{HandleShapeCreated} behaviour is a \texttt{cyclic} behaviour that processes such replies. Note that a \texttt{RequestNewShape} behaviour uses the \texttt{Shapes} ontology, and it can be activated by any type of agent. On the contrary, a \texttt{HandleShapeCreated} behaviour can be activated by \texttt{ShapeProvider} agents only because it needs to refer to the \texttt{myShape} property of the agent. Since a \texttt{HandleShapeCreated} behaviour is designed to work with \texttt{ShapeRequester} agents, it is assumed that it can implicitly use the \texttt{Shapes} ontology because the declaration of \texttt{ShapeRequester} agents ensures that the \texttt{Shapes} ontology is referenced.

\begin{figure}[t]
\begin{lstlisting}
agent ShapeRequester uses ontology Shapes
	property myShape as shape

	on create with providerName as text, x as double, y as double do
		activate behaviour HandleShapeCreated
		
		activate behaviour RequestNewShape with
			shapeProvider = aid(providerName),
			shapePosition = position(x, y)
\end{lstlisting}
\caption{Example of a \Jadescript agent with one property and one \texttt{on-create} handler that defines the procedure to be executed upon creation in terms of two behaviours.}
\label{fig:example_agent}
\end{figure}

\begin{figure}[ht]
\begin{lstlisting}
cyclic behaviour HandleShapeCreated for agent ShapeProvider
	on inform when content matches shapeCreated(s) do
		myShape of agent = s

one shot behaviour RequestNewShape uses ontology Shapes
	property myProvider as aid
	property myPosition as position

	on create with shapeProvider as aid, shapePosition as position do
		myProvider = shapeProvider
		myPosition = shapePosition

	do
		send request createShape(myPosition) to myProvider
\end{lstlisting}
\caption{The two behaviours used in the declaration of \texttt{ShapeRequester} agents shown in Figure~\ref{fig:example_agent}.}
\label{fig:example_behaviour}
\end{figure}

\section{A Support for Percept Handlers in Jadescript}
\label{sec:event_handling}
\Jadescript supports event-driven programming by means of event handlers included in agent and behaviour declarations. The current version of the language provides linguistic constructs for three categories of events: the transition to a new state in the lifecycle of an agent, the reception of a message from an agent, and the perception of an interesting event that occurred in the environment and that is described in terms of a new percept. Note that the support for percept handlers has been recently added to the language to serve the needs of robotic applications.

The \texttt{on-create} and \texttt{on-destroy} handlers are used to manage the events that represent changes in the lifecycle state of an agent, and they are declared in the scope of an agent declaration. The \texttt{on-create} handler is activated when an agent is started, and it supports an optional list of typed parameters to allow a set of arguments to be passed to an agent for its proper activation. The \texttt{on-destroy} handler is activated when an agent is requested to terminate, right before its termination procedure begins. 

Incoming messages directed to an agent are added to the internal message queue of the agent. They wait in the queue for \texttt{on-message} handlers to extract them and to perform the needed reactions. Such handlers are executed in the scope of a behaviour, and they only run when the behaviour is effectively scheduled for execution. The \texttt{on-message} handlers and the \texttt{on-percept} handlers (described below) can be provided with optional \texttt{when} expressions. A \texttt{when} expression defines a Boolean condition on the processed event and on the state of the behaviour that must be satisfied in order for the event handler to be considered applicable to the event. 
Any Boolean expression without side effects can be used as \texttt{when} expression. An important Boolean operator recently introduced in the language is the \texttt{matches} operator~\cite{jadescript:woa2019}, which allows to declaratively express the structure of the expected events by means of a pattern. Figure~\ref{fig:example_behaviour} shows an example of the use of a \texttt{when} expression and of the \texttt{matches} operator at line~$2$. Note that the \texttt{matches} operator can be used to declare local variables and to assign values to them by decomposing the structure of its left operand. For example, the local variable \texttt{s} in Figure~\ref{fig:example_behaviour} at line~$2$ is implicitly declared and assigned by the \texttt{matches} operator.

The \texttt{on-percept} handlers are a new addition to the language, and they represent the main mechanism provided by \Jadescript to receive percepts from the environment. Percepts are ontology schemas that extend the native predicate \texttt{percept}. Such a predicate features a single property, called \texttt{priority}, which is a natural number that represents the default priority assigned to a percept. Percepts with a higher property are treated with higher urgency. To achieve this, each agent maintains an internal priority queue that collects all the percepts received from the environment. Such a queue is then visited in decreasing order of priority, and appropriate \texttt{on-percept} handlers are possibly executed for each percept in the queue. Note that if a percept handler provides a \texttt{when} expression, such an expression can prevent the execution of the handler even if suitable percepts are in the priority queue.

The novel support for percept handlers is completed with a Java API to allow external event sources to decide when a percept should be provided to an agent. Such an API is based on the \emph{O2A} (\emph{Object-to-Agent}) interface~\cite{jade:book2007} that \JADE provides to pass Java objects to agents in a thread-safe way. Actually, the novel API to provide percepts to agents does not require explicit synchronization, and it assumes that it is used in a thread different from the threads of the agents that receive the percepts. The main methods offered by the API are \texttt{notifyEvent(Percept)} and \texttt{notifyEvent(Percept, int)}. Such methods are declared in the \texttt{JadescriptAgent} class, which is the Java class used as base class of all \Jadescript agents instead of the ordinary \texttt{Agent} class that \JADE provides. Both methods, when invoked, provide their first argument to the targeted agent via the O2A interface. The second method treats its second argument as a priority that is used to override the value of the \texttt{priority} property of percepts, which is treated just as a default priority.

Since the handling of percepts often requires agents to promptly react to events, the internal management of percepts, and the execution of related handlers, is integrated in the main loop of each agent to ensure that percept handlers are executed before behaviours. For this reason, the agent loop was changed with respect to the ordinary loop of \JADE agents, as shown in Algorithm~\ref{fig:agentloop}.

\begin{algorithm}[t]
\begin{algorithmic}
\Require $o2aQueue$, $priorityQueue$, $registeredHandlers$, $behaviourScheduler$
\Loop
	\While {\Not{} $o2aQueue$.isEmpty()}
		\State $p \gets o2aQueue$.peek()
		\If {$p$ \Is{} percept}
			\State $p \gets o2aQueue$.dequeue()
			\State $priorityQueue$.enqueue($p$)
		\EndIf
	\EndWhile
	\While {\Not{} $priorityQueue$.isEmpty()}
		\State $p \gets priorityQueue$.dequeue()
		\ForAll {$handlers$ \In{} $registeredHandlers$} 
			\If {$handler$.isApplicableTo($p$)}
				\State $handler$.execute($p$)
			\EndIf
		\EndFor
	\EndWhile
	\If {$behaviourScheduler$.hasNextBehaviour()}
		\State $b \gets behaviourScheduler$.getNextBehaviour()
		\State $b$.execute()
	\EndIf
\EndLoop
\end{algorithmic}
\caption{The internal loop of \Jadescript agents modified to support percept handlers.}
\label{fig:agentloop}
\end{algorithm}

\section{An Case-Study Application}
\label{sec:example}
Percept handlers have been recently added to \Jadescript to effectively support robotic applications like the case-study application described in this section, which was used to perform simple experiments. The discussed experiments consist of making a small robot perform autonomous explorations of an indoor environment avoiding obstacles. The explored environment is a known corridor, and an \emph{UWB} (\emph{Ultra-Wide Band}) localization infrastructure (e.g.,~\cite{electronics2019}) is installed to allow the robot to estimate its position on the floor of the corridor. Obstacles are cubes placed randomly in the corridor, and each face of each cube has a QR code attached. The mission of the robot is to explore the corridor avoiding obstacles to trace the approximate positions of all obstacles.

\begin{figure}[ht]
\centering
\includegraphics[height=0.5\textheight]{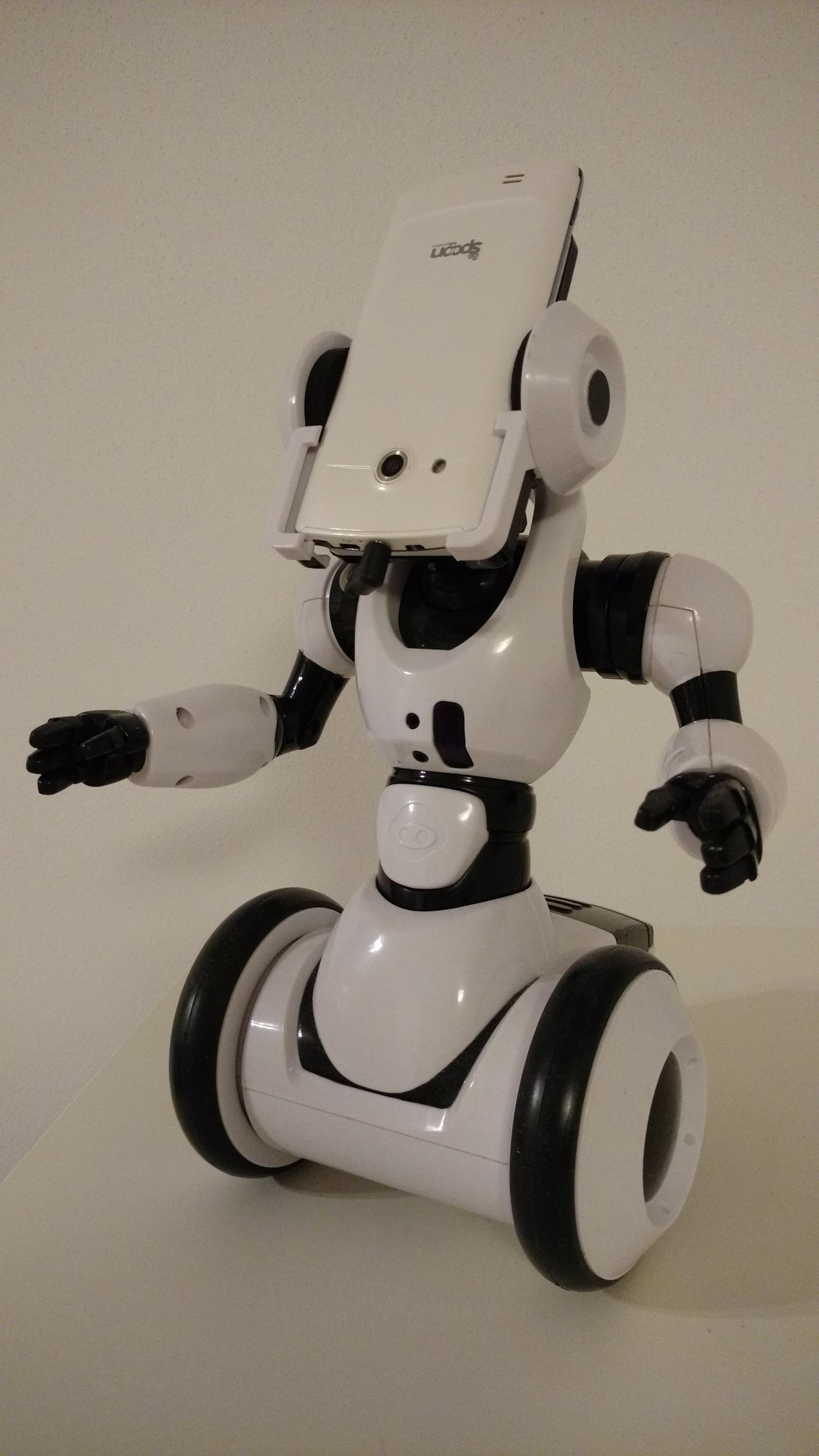}
\caption{The RoboMe used for experiments with the linked SpoonPhone installed in the head cradle and connected with an audio cable to send commands to the robot and to receive proximity alerts.}
\label{fig:robome_photo}
\end{figure}

RoboMe, a toy robot produced by WowWee (\url{http://www.wowwee.com}), was chosen to perform the described experiments. RoboMe is a simple robot because it has only one sensor, which is a front proximity sensor, and its actuators are just two slow wheels, a moving head cradle, and a loudspeaker. Besides being so simple, RoboMe was chosen because it can be easily connected to a smartphone that can be set firmly on its head cradle. The attached smartphone provides the robot with much more sensors, and it can easily control the robot with a simple API. The smartphone is linked to the robot using an audio cable that connects to the headset port of the smartphone, and the exchange of information between the robot and the attached smartphone is based on sounds. The robot reacts to a fixed set of sounds, and each sound is associated with a particular action, which include stop, move forward, move backward, turn left, turn right, lower the head cradle, and raise the head cradle. Similarly, the robot signals the presence of front obstacles using a specific sound that is interpreted by the smartphone.

The smartphone used for the discussed experiment is a SpoonPhone, which is a smartphone produced by BeSpoon~(\url{http://www.bespoon.com}). The main feature of SpoonPhones is that they integrate hardware and software modules needed to perform accurate positioning with respect to an UWB infrastructure. SpoonPhones have already been used to perform experiments on indoor positioning, and they ensured a position accuracy of less than $1$~m (e.g.,~\cite{electronics2020}), which is considered sufficient for targeted applications~\cite{amuse:paams2016}. In particular, the support for the UWB sensors that SpoonPhones provide have already been integrated in the localization add-on module~\cite{loc:paams2016} for \JADE to perform the mentioned experiments on indoor positioning. In addition, SpoonPhones are ordinary Android smartphones, and the SpoonPhone used for experiments featured an accelerometer, a gyroscope, a compass, and a front camera. Figure~\ref{fig:robome_photo} shows a picture of the RoboMe used for experiments with the linked SpoonPhone installed in the head cradle and connected with an audio cable to send commands to the robot and to receive proximity alerts. Note that the arms and the hands of the robot are not motorized.

In order to perform the discussed experiments and make the robot autonomously explore the environment, a dedicated Android application for the SpoonPhone was developed mostly in \Jadescript. WowWee provides a Java API for RoboMe (\url{http://github.com/WowWeeLabs/RoboMe-Android-SDK}) that allows Android applications to send commands to the robot and to receive proximity alerts. Therefore, the developed Android application bridges the RoboMe API with a \Jadescript agent hosted on a \JADE container deployed using the \JADE add-on for Android~\cite{jade:android}. The RoboMe API has nonblocking methods, and the commands to the robot can be safely sent from \Jadescript behaviours because they do not block the agent loop. In addition, the Java part of the application uses the localization add-on module for \JADE to estimate the position of the robot in the indoor environment and to constantly inform the agent of its position using suitable percepts. Finally, the Java part of the application uses ordinary APIs to read QR codes and to gather meaningful information from the accelerometer, the gyroscope, and the compass. The Java part of the application is executed in a different thread with respect to the thread of the agent. The Java API discussed in the previous section is used to provide the agent with the percepts gathered using the mentioned event sources in a thread-safe way. 


The structure of the \Jadescript agent can be briefly summarized as follows. A \texttt{RoboMe} agent uses a \texttt{HandlePercepts} behaviour to handle percepts generated by the Java part of the Android application. Such percepts are defined in the \texttt{RoboMeApplication} ontology, which is shared among the agent, its behaviour, and the Java part of the Android application. Figure~\ref{fig:robome_onto} shows the \Jadescript source code of the ontology. The ontology includes four types of percepts associated with different priorities. The \texttt{falling} percepts are generated when the sensors of the smartphone (accelerometer and gyroscope) can be used to alert the agent that the smartphone is falling. Note that such percepts have the highest priority. The \texttt{obstacle} percepts are generated when the proximity sensor of the robot notifies the presence of a front obstacle. Note that such percepts have a high priority, but they do not have the highest priority. The \texttt{qrCodeRead} percepts are generated when the images from the front camera of the smartphone can be used to detect a QR code. Finally, the \texttt{position} percepts are generated by the localization add-on module for \JADE at constant rate using the ranging information acquired from the UWB sensor of the SpoonPhone. Note that the \texttt{falling} percepts and the \texttt{obstacle} percepts are defined as propositions because they simply notify that an event has recently occurred, but they do not carry other information. On the contrary, the \texttt{qrCodeRead} percepts and the \texttt{position} percepts carry meaningful information, and therefore they are declared as predicates.

\begin{figure}[tt]
\begin{lstlisting}
module it.unipr.ailab.robome

ontology RoboMeApplication
	proposition falling extends percept with priority = 10
	
	proposition obstacle extends percept with priority = 8

	predicate qrCodeRead(qrCode as string) extends percept 
		with priority = 4

	predicate position(x as double, y as double) extends percept 
		with priority = 2
\end{lstlisting}
\caption{The \Jadescript ontology used in the case-study application to declare the types of percepts that the Java part of the application for the SpoonPhone can generate using relevant information from sensors.}
\label{fig:robome_onto}
\end{figure}

\begin{figure}[ht]
\begin{lstlisting}
module it.unipr.ailab.robome

agent RoboMe uses ontology RoboMeApplication
	property roboMeInterface as any

	property worldMap = {}
	
	property currentPosition = position(0, 0)
	
	on create with interface as any do
		roboMeInterface = interface
		
		invoke "startListening" on roboMeInterface 
		
		activate behaviour HandlePercepts

		do moveForward

	procedure stop do
		invoke "callStop" on roboMeInterface
			
	procedure moveForward do
		invoke "callMoveForward" on roboMeInterface

	procedure turnLeft do
		invoke "callTurnLeft" on roboMeInterface
		
	# other procedures that use roboMeInterface were removed
	# for the sake of brevity
\end{lstlisting}
\caption{Some of the most relevant parts of the \Jadescript source code of the agent developed for the case-study application.}
\label{fig:robome}
\end{figure}

Some of the most relevant parts of the \Jadescript source code of the agent developed for the case-study application are shown in Figure~\ref{fig:robome}. The mission of the agent is to move the robot in the forward direction while trying to capture all visible QR codes, and when an obstacle is detected, to change its direction to avoid the obstacle. The \texttt{HandlePercepts} behaviour, which is partially shown in Figure~\ref{fig:robome_behaviour}, is used to process percepts and to avoid obstacles, while tracking the approximate positions of visible QR codes. Note that the Java part of the Android application is responsible for creating the \JADE container, for creating a Java object to allow the agent to interface the actuators of the robot, and to create a \texttt{RoboMe} agent. When the \texttt{RoboMe} agent is created, it receives the Java object that it can use to command actuators. The Java object, which is referenced as \texttt{roboMeInterface} in the source code of the agent, provides a Java method for each available command to actuators, and the agent uses such methods directly with the \texttt{invoke} keyword. \Jadescript offers the \texttt{any} data type for opaque Java objects, and it provides the \texttt{invoke} keyword to invoke methods on Java objects, so that \Jadescript agents can take full benefit of libraries and frameworks available to the hosting Java virtual machine. Finally, note that the first method invoked on the \texttt{roboMeInterface} is \texttt{startListening}, which immediately activates the delivery of percepts to the agent from all supported event sources.

Some of the most relevant parts of the \Jadescript source code of the \texttt{HandlePercepts} behaviour are shown in Figure~\ref{fig:robome_behaviour}. In particular, Figure~\ref{fig:robome_behaviour} shows the procedures that are performed when \texttt{obstacle} percepts, \texttt{qrCodeRead} percepts, and \texttt{position} percepts are received. When an \texttt{obstacle} percept is received, the robot is immediately stopped, then it is requested to turn left, and it is finally requested to restart. When a \texttt{qrCodeRead} percept is received, the text of the read QR code is associated with the current position of the robot in the map of the world. Finally, when a \texttt{position} percept is received, the property of the agent that stores the current position of the robot is updated.

\begin{figure}[t]
\begin{lstlisting}	
module it.unipr.ailab.robome

cyclic behaviour HandlePercepts for agent RoboMe
	on percept when content matches obstacle do
		do stop
		do turnLeft
		do moveForward

	on percept when content matches qrCodeRead(qrCode) do
		worldMap[currentPosition] = qrCode

	on percept when content matches position(x, y) do
		currentPosition = position(x, y)
		
	# other percept handlers were removed for the sake of brevity
\end{lstlisting}
\caption{Some of the most relevant parts of the \Jadescript source code of the \texttt{HandlePercepts} behaviour used by \texttt{RoboMe} agents.}
\label{fig:robome_behaviour}
\end{figure}

\section{Conclusion}
\label{sec:conclusions}
This paper presented exploratory experiments on the use of an AOP language to program autonomous robots. First, the paper briefly recalled some of the major features of \Jadescript. Then, it presented the novel support for percept handlers that has been recently introduced in the language to serve the needs of robotic applications. Finally, it briefly described a simple case-study application designed to validate the possibility of using \Jadescript to program autonomous robots. Even for such a simple application, the support for event handling in the language was needed so urgently that the development plans for the language were changed to anticipate it. The design choices that form the basis of the language ensured that such a new feature could be incorporated easily and coherently with the rest of the language, and the experiments presented in this paper suggest that now \Jadescript can be used to better support the event-driven programming of autonomous robots.

The management of the priorities of percepts, and of percepts with respect to behaviours, which is currently solved with a priority queue, calls for further investigations. The current implementation assumes that percepts are more important than messages.
Hence, messages are processed only when all percepts have already been handled, which might not be sufficient in some situations. 
In addition, the current approach to sense the environment and to control actuators should be further questioned. A possible improvement of the current approach could be the introduction in the language of the explicit support for sensors and actuators to match the recent introduction of the support for percepts.

\bibliographystyle{eptcs}
\bibliography{bibliography}

\end{document}